**Processing Issues in Top-Down Approaches to Quantum Computer Development in Silicon**

S.-J. Park[1,] A. Persaud[1], J. A. Liddle[1], J. Nilsson[2], J. Bokor[1,3], D. H. Schneider[3], I. Rangelow[4] and T. Schenkel[1]
[1]Lawrence Berkeley National Laboratory, Berkeley, CA, USA
[2]Lawrence Livermore National Laboratory, Livermore, CA, USA
[3]Department of EECS, University of California, Berkeley, CA, USA
[4]Institute of Microstructure Technologies and Analytics, University of Kassel, Germany

jaliddle@lbl.gov

## ABSTRACT

We describe critical processing issues in our development of single atom devices for solid-state quantum information processing. Integration of single $^{31}P$ atoms with control gates and single electron transistor (SET) readout structures is addressed in a silicon-based approach. Results on electrical activation of low energy (15 keV) P implants in silicon show a strong dose effect on the electrical activation fractions. We identify dopant segregation to the $SiO_2$/Si interface during rapid thermal annealing as a dopant loss channel and discuss measures of minimizing it. Silicon nanowire SET pairs with nanowire width of 10 to 20 nm are formed by electron-beam lithography in SOI. We present first results from Coulomb blockade experiments and discuss issues of control gate integration for sub-40 nm gate pitch levels.

## INTRODUCTION

The potential of quantum systems to process information very efficiently has been under intensive investigation since the development of suitable algorithms by Shor[1] and Grover.[2] There are a number of techniques for creating suitable systems such as NMR,[3] ion trapping and a variety of solid-state approaches.[4, 5] Although NMR and ion trap methods have successfully demonstrated the fundamental concepts, it is unlikely that they can be scaled effectively to produce an integrated system of qubits to the point where it becomes computationally interesting (i.e. > 1000 qubits). For this reason we have chosen to examine in detail a solid state scheme because such an approach holds out the promise of full scalability. Further, we have chosen to use a Si-based architecture because it can, in principle, be integrated with more conventional electronic devices and because the intensive study of Si and Si-processing makes it more likely that we can achieve the necessary atomic scale control of structural and electronic material properties.

Electron and nuclear spins of $^{31}P$ atoms in silicon are promising candidates for the realization of a scalable solid-state quantum computer architecture (Figure 1).[6,7,8] Three components are required in order to fabricate a test device: 1) an array of single, activated $^{31}P$ atoms, 2) single spin readout devices (e.g. single electron transistors (SETs)[9]), and 3) integrated control gates.[10]

It is crucial to perform experimental tests of the key tasks that need to be addressed for basic qubit scheme validation: single spin manipulation, single spin detection, coherent electron transfer, and two qubit operations. Our development of an integrated two-qubit device is thus geared to provide access to the physics of the $^{31}P$ qubit in silicon.

In this paper we will describe our development of single ion implantation technology, our current process flow, including the initial results of our Si SET effort, and discuss the challenges involved in completing the device with integrated control gates.

## SINGLE ION IMPLANTATION

Three criteria must be met for the successful formation of a single atom array of $^{31}P$ qubits: 1) individual atoms must be delivered to the substrate, 2) they must be electrically active and 3) they must be in a known and stable location.

Single ion implantation with low-energy (<20 keV), highly-charged ions offers a path to the formation of single $^{31}P$ atom arrays.[11] Figure 2 a) shows an illustration of our single ion implantation system – individual ions pass through a small aperture fabricated at the end of a conventional SPM probe tip.[12] The tip can be positioned precisely with respect to substrate topography and the implantation of a single ion is detected by the burst of secondary electrons emitted when it is incident on the surface. Figure 2 b) shows an example of pulse height distributions from detection of secondary electron bursts from low energy xenon ions. The high charge-state of the ion ensures that a sufficient number of secondaries are emitted with each impact to guarantee an unambiguous detection signal.[11] The small aperture size and its precise location ensure that each atom is precisely placed. Apertures with diameters as small as 4.3 nm have been formed by focused ion beam drilling in combination with local thin film deposition.[12]

A qubit spacing of 20 nm, equivalent to about 10 effective Bohr radii of the bound donor electron of a P atom in a silicon matrix, was envisioned in the original Kane proposal.[6] This spacing allows two qubit operations through exchange interaction mediated by direct wave function overlap.[13] $^{31}P$ atom spacings of ~100 nm are possible when spin-coherent electron shuttling can be developed as a means for two qubit interactions. The equivalent doses for qubit arrays are $10^{10} – 2\times10^{11}$ $^{31}P/cm^2$. In single ion implantation, ions are implanted to a depth of about 10 to 20 nm. An implant energy of 10 – 20 keV is favorable since it allows implantation through a pre-formed gate dielectric with minimal dopant loss in the dielectric layer. Also, the single ion positioning uncertainty from range straggling is kept below 10 nm.

Following implantation, wafers have to be annealed for damage repair and to ensure electrical activation of the dopants. Diffusion and segregation of dopants during rapid thermal annealing have to be minimized, while activation efficiencies close to 100% are mandatory for devices whose functionality is based on control of electron spin states in single atoms.

In Figure 3 a), we show spreading resistance analysis (SRA) data from a 60 keV, $10^{11}$ $cm^{-2}$ implant of P in silicon. Implanting at 60 keV places the peak of the dopant distribution at a depth of 30 nm and allows us to delineate a segregation effect with SRA. Data in the figure compare the post annealing profiles for $SiO_2$ and $Si_3N_4$ surface layers. For $SiO_2$, we observe strong segregation towards the interface, while for $Si_3N_4$, we do not. The activation ratios, i.e., the ratio of measured carrier concentration to implanted dose, are 80% for both samples. However, the strong segregation effect for the oxide limits activation ratios for lower energy implants. The dose dependence of P activation in silicon is shown in Figure 3 b) for 15 keV implants. Very low electrical activation fractions for low dose implants can be attributed to dopant loss at the interface. $SiO_2$ injects interstitials during rapid thermal annealing that aid in the segregation of P atoms towards the interface. P atoms can replace under-coordinated silicon atoms at the interface. The P remains then in a neutral state and is not electrically active.[14] The observed dopant loss mechanism is dose dependent, and we are currently investigating co-implantation of silicon ions to change the interstitial vacancy balance, together with aggressive pre-implant anneals to further enhance activation ratios in the low dose, low energy implant regime.

## PROCESS INTEGRATION

Figure 4 outlines the process flow, while Figure 5 shows an SEM image of a Si-nanowire SET pair in silicon-on-insulator (SOI). Here, the nanowire thickness is 30 nm, and the wire region is undoped. Gate, source and drain regions were highly n-type doped. Figure 5 a) shows initial results from electrical characterization of devices at 4.2° K.

Coulomb blockade effects have been observed in silicon nanostructures for many years.[15] In our application of Si-SETs as sensitive electrometers in spin dependent charge measurements,[9] operation temperatures will be limited to <1° K by low energy splittings of spin states in external magnetic fields. There is also a need to minimize decoherence from coupling of qubits to environmental noise sources in the silicon matrix. Consequently, the size requirements for our SETs are relaxed, compared to developments of SETs for higher temperature applications. Coulomb blockade is likely to result from defects and local inhomogeneities in the nanowire and the interface of the nanowire to the highly n-type doped source and drain leads. While Coulomb oscillations of regular period are desirable, stability of device characteristics is of even greater importance for SET sensors. We are now investigating the latter in systematic studies.

An important criticism of solid-state approaches to quantum computer development is that all devices components will have minute differences,[16] which will result in a need for diligent calibration of components. SOI allows a high degree of control over process parameters, and is compatible with high temperature processes. The latter is important for reduction of defect densities. Defects stable on the time scale of $10^4$ qubit manipulations (tens of milliseconds) might also be permissible.

## CONTROL GATES

Local control of electron wave functions is at the heart of all atom based solid-state qubit proposals. This local control can be achieved with electrical gates, which tune single electrons into resonance with external magnetic fields for single qubit operations. For two qubit operations, a control gate can act on the extent of electron wave function overlap between neighboring atoms. Gate width requirements depend on the qubit spacing. In the original Kane proposal, two A gates, and one J gate are required with a qubit spacing of 20 nm. An equivalent gate pitch of 10 nm (5 nm gate width, and 5 nm spacing) is beyond current state-of-the-art for dense patterns. While isolated lines with widths of only a few nm have been formed for many years, the key problem is the development of dense pattern capabilities at a 10 nm gate pitch. An additional boundary condition is that gates have to be aligned to within a few nm to the single atom arrays. Gate integration at this level is a crucial issue both for top-down (with single ion implantation) and bottom-up (with scanning tunneling microscopy)[17,18] based device fabrication strategies. Gateless, all-optical control of atom qubits has recently been proposed, but issues of multi-qubit entanglement remain to be addressed.[19]

Alternatives to a 10 nm gate pitch require strategies to increase the qubit spacing. Use of silicon germanium hetero-structures has been suggested by Virjen et al.[20] The resulting g-factor engineering results in effective Bohr radii of about an order of magnitude in excess of values for a bare silicon matrix. Design of the hetero-structures comes with its own challenges. An alternative is to shuttle electrons between P atoms to mediate two-qubit operations. This requires encoding of a logical qubit in the electron and nuclear spin of a P atom. It also requires design of silicon quantum well structures that allow reliable movement of a single electron over distances of hundreds of lattice spacings, without losing the electron and without modifying its carefully prepared spin state (orientation and phase). Clearly, spin coherent single electron transport is a crucial experimental challenge in silicon based quantum computer development. An important recent result is that spin coherence times have been shown to depend on P atom spacing: for larger P atom separations (>100 nm), decoherence times of tens of milliseconds have been found in bulk measurements.[21]

## CONCLUSIONS

In this article we address several critical issues in the development of a two $^{31}P$ qubit device in silicon that allows testing of basic operations, and that can be used to validate basic quantum information processing schemes. Single ion implantation aligned with a scanning probe is a path to formation of nanometer scale single P atom arrays in silicon. Single ion registration is achieved through detection of secondary electron bursts from highly charged ion impacts. Achieving essentially full electrical activation of a low energy, low dose P implant is crucial for device development. We show an effect of dopant segregation towards the $SiO_2$/Si interface at the $10^{11}$ $cm^{-2}$ dose level. Measures to eliminate this effect currently under investigation are: annealing under nitridation conditions, aggressive hydrogen anneals before implantation, and co-implantation of silicon.

Silicon nanowire SET pairs were formed in SOI with line width of 10 to 20 nm. We discuss lithography and pattern transfer process issues and present first results from Coulomb blockade studies at liquid helium temperatures. Design and fabrication of control gates is another critical issue, where two qubit operation schemes and lithographically achievable gate widths have to be balanced.

While significant technical challenges remain to be resolved, we present a workable processing path towards achieving fabrication of all silicon two qubit test structures in the near future.

## ACKNOWLEDGMENTS:

This work was supported by the NSA and ARDA under ARO contract number MOD707501, and by the U. S. DOE under contract No. DE-AC03-76SF00098. Work at LLNL was performed under the auspices of the U. S. DOE under contract No. W-7405-ENG-48.

## FIGURE CAPTIONS

Figure 1. Conceptual layout of a two-qubit test structure.

Figure 2 a) Illustration of the single ion implantation scheme where individual highly charged dopant ions are implanted aligned by scanning probe microscopy (SPM) to sample features. Following one ion strike, the beam is blocked, and the sample moved to the next implant position. Apertures are formed by FIB drilling of ~100 nm holes and size reduction by local deposition of a Pt film. Components are not to scale. 1) SPM cantilever, 2) Hollow SPM tip with 30 nm wide aperture, 3) SPM image of a Silicon nanowire SET pair, 4) highly-charged dopant ion impacts, 5) secondary electron emission and detection.[12] b) Example of pulse height distributions from detection of secondary electrons in the single ion implanter following the impact of low energy ions on silicon. The kinetic energies of $Xe^{32+}$ ions were 11 keV (gray) and 7 keV (black). Potential electron emission yields increase for lower impact energies enabling efficient detection of very low energy ions

Figure 3 a) Carrier concentration as a function of depth from spreading resistance analysis of P implanted silicon (60 keV, $10^{11}$ cm$^{-2}$). Black: 7 nm silicon nitride layer, gray: 5 nm thermal $SiO_2$ layer. b) Activation ratios determined by spreading resistance analysis as a function of ion dose. The corresponding dose for a Kane-type solid-state quantum computer is $10^{10}$ – $2\times10^{11}$cm$^{-2}$. All samples were annealed for 10 s at 1000° C in a $N_2$ atmosphere.

Figure 4. Process flow. Alignment marks are created in a SiGe layer and are used for both optical and electron beam lithography. The SET islands and source and drain are defined in a thin (< 30 nm) SOI layer. The island regions are masked during the implantation step that creates the source and drain. Oxide is deposited over the SET structure. Lateral epitaxial overgrowth (LEO) is used to create a high-quality $^{28}$Si layer and a gate oxide is grown upon it. The $^{31}$P ions are implanted and activated with an anneal. The control gates are then fabricated. Finally a series of conventional optical lithography steps create the contacts to the small-scale features.

Figure 5 a) SEM of a Si SET-pair structure in silicon-on-insulator. The silicon nanowire is undoped. The SOI thickness is 30 nm on 400 nm buried oxide (BOX). Source, drain and gate leads are highly n-type doped. b) Coulomb blockade oscillations in a silicon nanowire SET at 4.2° K.

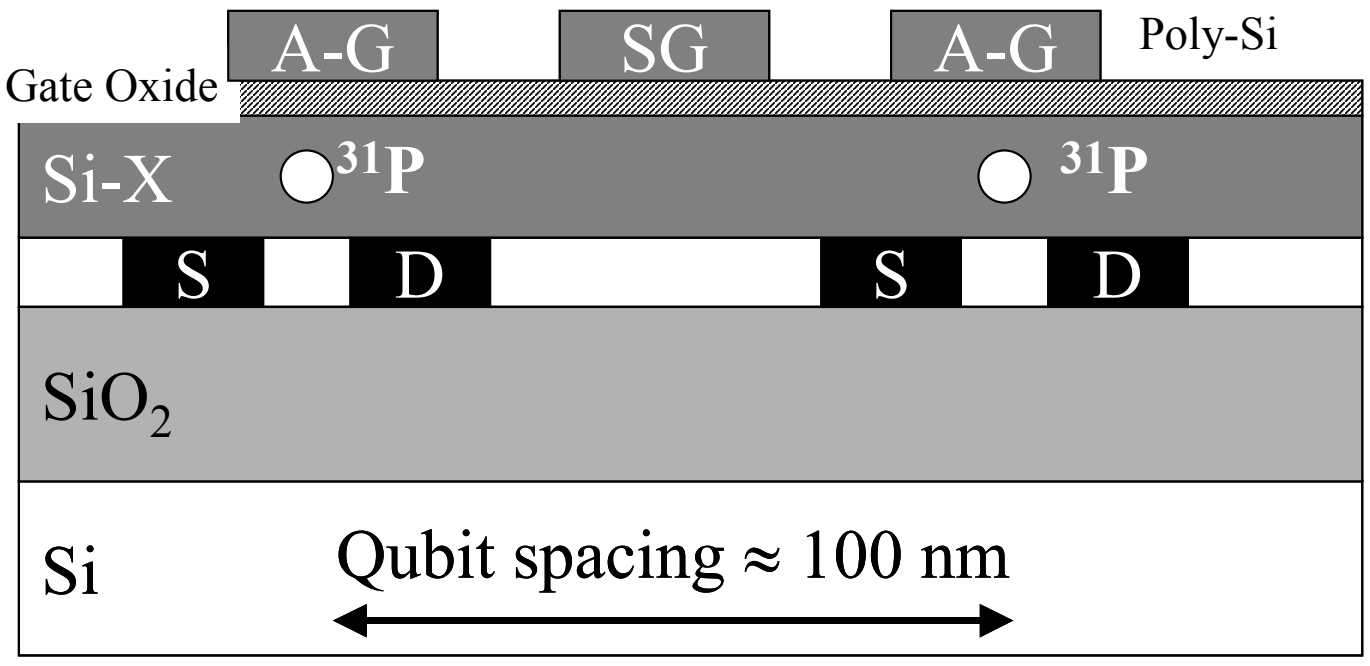

Figure 1.

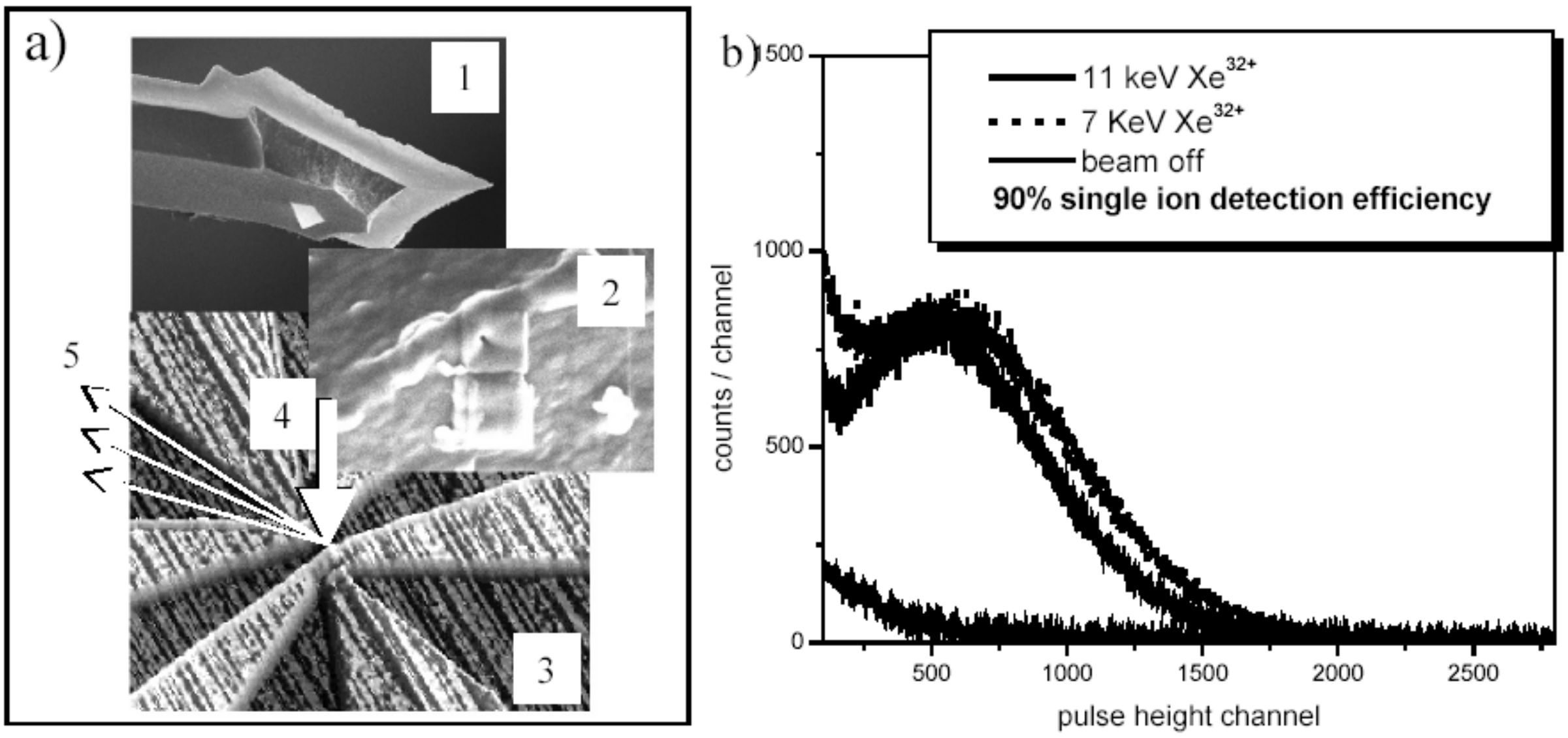

Figure 2.

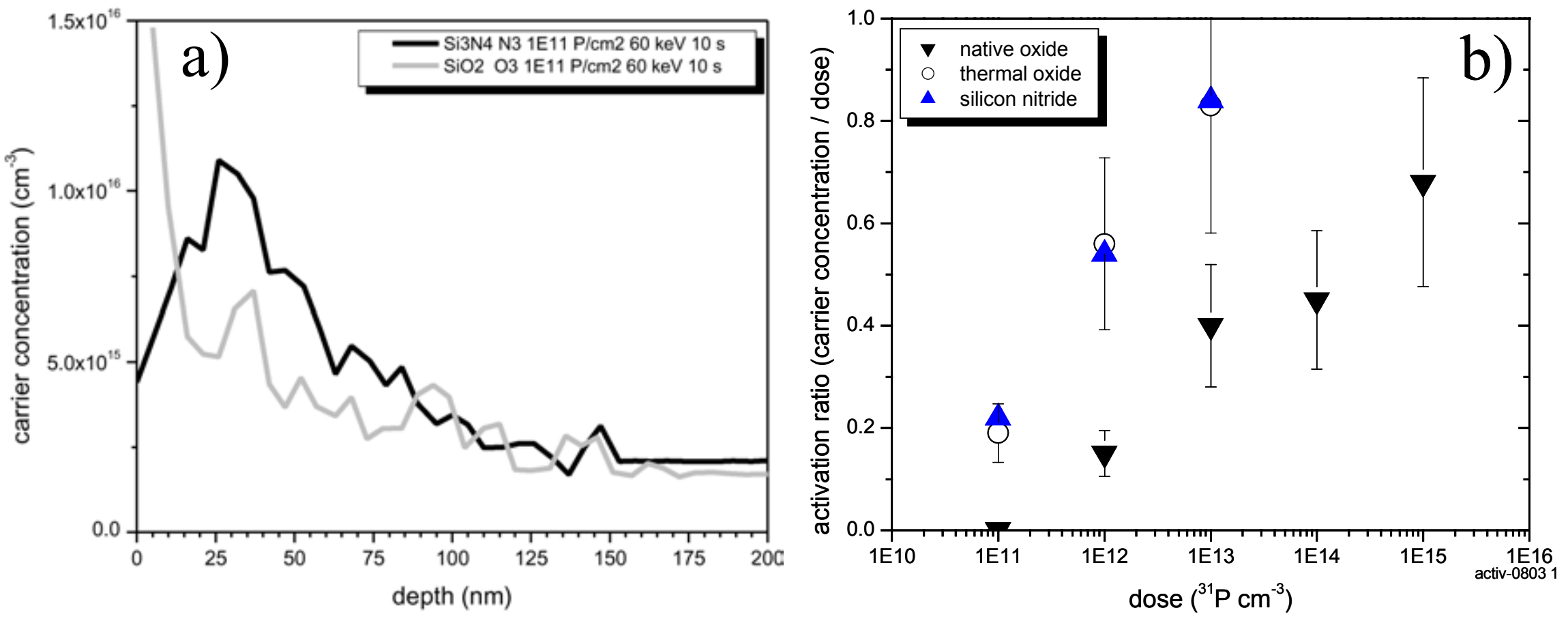

Figure 3.

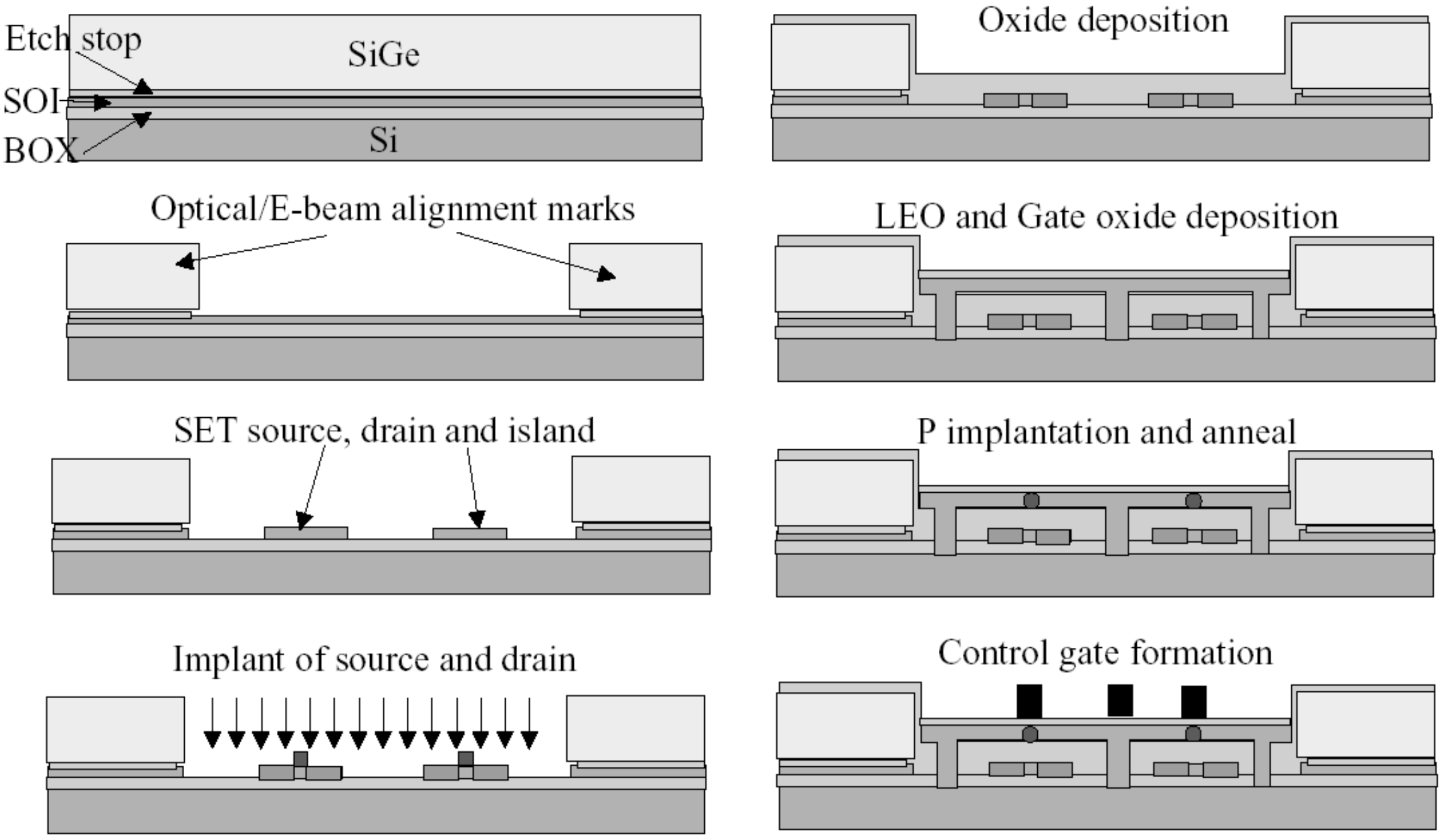

Figure 4.

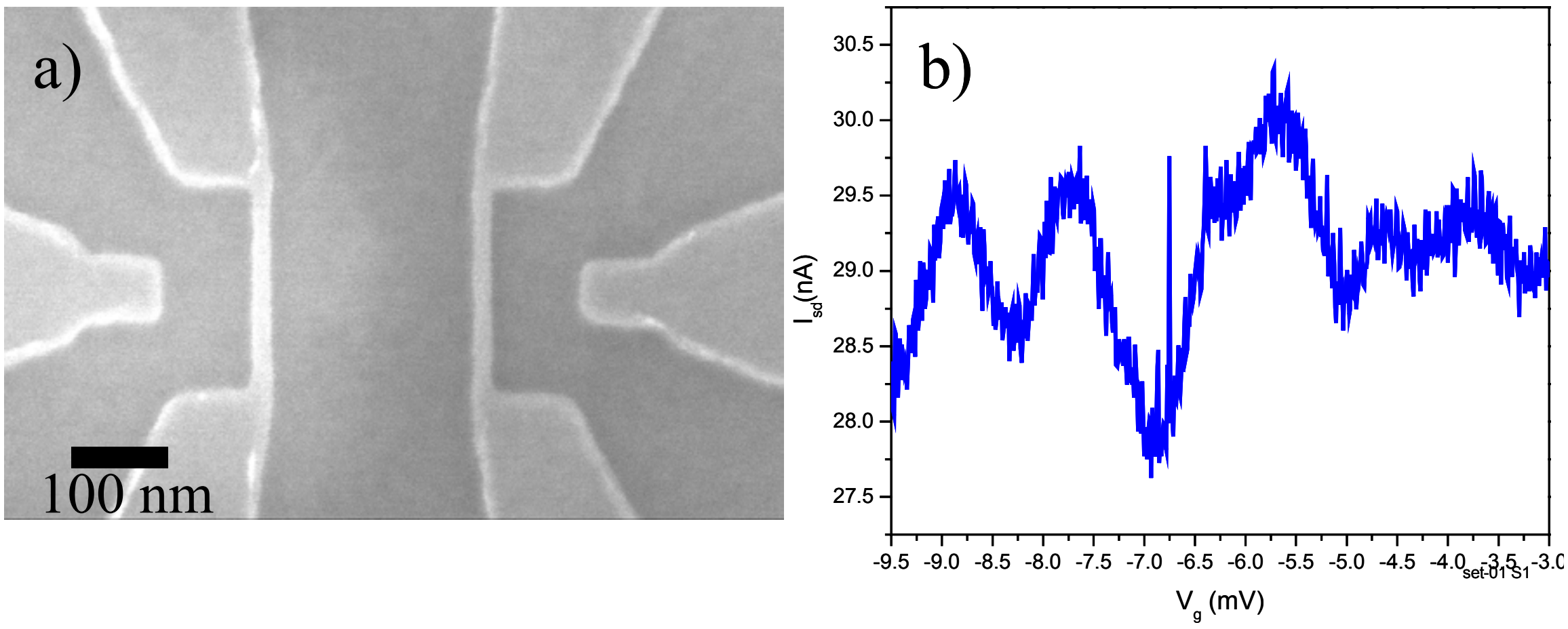

Figure 5.